# Photoluminescence and photochemistry of the $V_B^-$ defect in hexagonal boron nitride

Jeffrey R. Reimers,[1,2] Jun Shen,[3] Mehran Kianinia,[2] Carlo Bradac,[2] Igor Aharonovich,[2] Michael J. Ford,[2] and Piotr Piecuch[3,4]

[1] *International Centre for Quantum and Molecular Structures and Department of Physics, Shanghai University, Shanghai 200444, China.*
[2] *University of Technology Sydney, School of Mathematical and Physical Sciences, Ultimo, New South Wales 2007, Australia.*
[3] *Department of Chemistry, Michigan State University, East Lansing, Michigan 48824, United States of America.*
[4] *Department of Physics and Astronomy, Michigan State University, East Lansing, Michigan 48824, United States of America.*

Email: jeffrey.reimers@uts.edu.au; jun@chemistry.msu.edu; mike.ford@uts.edu.au; piecuch@chemistry.msu.edu

**Abstract**

Extensive photochemical and spectroscopic properties of the $V_B^-$ defect in hexagonal boron nitride are calculated, concluding that the observed photoemission associated with recently observed optically-detected magnetic resonance is most likely of $(1)^3E'' \rightarrow (1)^3A_2'$ origin. Rapid intersystem crossing from the defect's triplet to singlet manifolds explains the observed short excited-state lifetime and very low quantum yield. New experimental results reveal smaller intrinsic spectral bandwidths than previously recognized, interpreted in terms spectral narrowing and zero-phonon-line shifting induced by the Jahn-Teller effect. Different types of computational methods are applied to map out the complex triplet and singlet defect manifolds, including the doubly ionised formulation of the equation-of-motion coupled-cluster theory that is designed to deal with the open-shell nature of defect states, and mixed quantum-mechanics/molecular-mechanics schemes enabling 5763-atom simulations. Two other energetically feasible spectral assignments from amongst the singlet and triplet manifolds are considered, but ruled out based on inappropriate photochemical properties.

Hexagonal boron nitride (h-BN) has become of great interest following the 2016 discovery of single-photon emission from atom-like defects in the material[1-4]. Of significance is the recent observation of optically detected magnetic resonance (ODMR) associated with (at least) two types of h-BN defects[5,6]. This could enhance the use of h-BN defects in nanophotonic applications[7-10]. Much effort has been devoted to determining the chemical nature of different defects[11], including broad-based scans of possibilities[4] and detailed studies[12-14]. Prior to the detection of ODMR, no defect displaying photoluminescence had been assigned, only defects with observed magnetic properties[15,16]; defects exhibiting ODMR facilitate measurement of both the magnetic and photoluminescence properties. For one defect displaying ODMR[5], comparison of the observed magnetic properties to expectations[12] strongly suggested that the ODMR arises from the $V_B^-$ defect (a charged boron vacancy), an interpretation that was quickly supported quantitatively.[17,18] Calculations have predicted that $V_B^-$ has a triplet ground state[12,19], with a variety of low-energy triplet excited states predicted, within likely error limits, to have energies consistent with the observed photoemission energy[17,18].

The low-energy triplet manifold of $V_B^-$ is very complex, and subtle changes in its capture by different computational methods can have profound consequences on the predicted spectra. A significant issue is that spectra predicted for the lowest-energy transitions are very broad and inconsistent with the experimental observations.[18] Further, the observed spectrum corresponds to an ensemble of emitters at high temperature and could therefore be considerably broader than that for a single emitter, as modelled in the calculations. A key unexplained property of the observed emission is that it is very weak, indicating that some, currently unknown, photochemical process(es) must act to reduce the quantum yield upon photoexcitation. Intersystem crossing to the singlet manifold is a possible mechanism for this reduction in quantum yield, raising the possibility that the observed emission occurs from within the singlet manifold. We also note that previous estimates of spectral bandwidth based on calculations have assumed that the spectra obey simple relationships such as those expected based on Huang-Rhys factors depicting transitions between non-interacting, non-degenerate electronic states. We remark that, in contrast, $V_B^-$ is intrinsically a Jahn-Teller system involving many doubly degenerate electronic states, with the lowest-energy excited states also involving unmapped avoided crossings in the Franck-Condon region[18]. To understand how $V_B^-$ sustains ODMR, in this work we address a series of outstanding challenges[18]:

- Improved measurements of spectral bandshape are obtained, including measurements of its temperature dependence.
- Accurate transition-energy calculations are obtained, with boundable error estimates, for both the singlet and triplet manifolds.
- Accurate calculations of photoluminescence band shapes and intensities are performed, taking into effect long-range acoustic phonons, the Jahn-Teller effect, and the Herzberg-Teller effect.
- Useful estimates of many photochemical reaction rates are obtained, taking into reaction account transition states, as well as the influence of Jahn-Teller and other conical intersections.
- Verification of the consistency of our computational predictions against experimental spectroscopy measurement from samples showing the ODMR photo-dynamics, focusing on transition energies, bandwidths, emission lifetimes, quantum yields, and their temperature dependencies.

The $V_B^-$ defect has intrinsic $D_{3h}$ local point-group symmetry, when isolated within the bulk of an h-BN sheet[12,19-21]. All modelling reported herein is applied to the model compounds shown in Figure 1 that can all display this symmetry. The observed ODMR properties are suggestive of $D_{3h}$ symmetry, although some observed properties require slight symmetry lowering.[5] Note that $V_B^-$ defects are created by neutron/ion irradiation of h-BN, which can lead to damage in the crystalline structure and thus to lowering of symmetry. To date, symmetry lowering induced by local strain,[17] or by the defect being located at the edge[11,22,23] of a h-BN sheet[18], have been considered. Whereas some calculated properties may change dramatically based on these variations, the basic spectral properties are insensitive to such effects—which at any rate tend to generally broaden the spectra rather than to narrow them[18].

We have previously considered other defects in h-BN, seeking calibrated computational quantum chemistry methods and the conditions in which these methods can deliver accurate results.[24] We identified a number of issues, summarised in Table 1, that must be addressed in any computational work to model reliably the defect and its properties. A critical aspect is that many of the states of interest are open-shell in nature. We remark that none of the widely-used density functional theory (DFT) and *ab initio* wavefunction methods in traditional software packages are capable of accurately characterizing all defect properties of interest. The development of a reliable computational protocol for the prediction of defect properties should thus stem from the thorough, comparative analysis of a wider variety of feasible approaches than that considered in our previous work[24], including newer generations of *ab initio* quantum chemistry methods.

The electronic-structure computational methods used are listed in the Methods section, with their strengths and limits of applicability being discussed in the Supporting Information (SI) Section S1 and summarised in Table 2. Most methods are well known, but we add one new approach that is particularly well-suited to address defect spectroscopy and the singlet manifold of $V_B^-$ specifically. This is a coupled-cluster (CC) method utilising the equation-of-motion (EOM) formalism (EOMCC) in its double-ionisation-potential (DIP) version,[25-30] which involves removing two electrons from a reference wavefunction in order to create the electronic states of interest. The DIP-EOMCC approach provides an accurate description of open-shell character, assuming that the reference wavefunction is predominantly closed-shell in nature. We find it to be more suitable than traditional ways of treating open-shell systems such as complete-active-space self-consistent field (CASSCF)[31-33] and its extension, multi-reference configuration interaction (MRCI)[34], for it is more conducive to expansion towards the exact answer and hence the estimation of likely error bounds. For the triplet manifold, its ground state appears mostly single-reference in nature and so we apply traditional particle conserving EOMCC singles and doubles[35] (EOMCCSD) and time-dependent density-functional theory[36] (TDDFT) approaches. Another significant aspect of the Methods, applied to model 10-ring and larger compounds, is the use of mixed quantum-mechanics/molecular-mechanics (QM/MM) approaches utilizing, in the MM part, an AMBER[37] force field fitted to reproduce DFT-calculated properties of h-BN[38].

Simulation of spectra of defects in h-BN is usually made based on the Huang-Rhys approximation[39] or its variants. The basic model involves five core approximations, any or all of which are likely to fail when applied to defect spectroscopy[11]. Notably, this approach is inapplicable at conical intersections. General simulation codes pertinent to the Jahn-Teller effect are not available. We herein apply standard methods to model systems developed to mimic the key features of the triplet and singlet manifolds revealed by the electronic structure calculations. A key quantity of interest is the emission reorganisation energy $\lambda^E$, which we determine through the measurement and interpretation of experimental photoluminescence spectra. In semiclassical models such as the Huang-Rhys model, $\lambda^E$ equals the difference in energy on the final electronic state after a vertical transition (i.e., transition without change in defect geometry induced by interactions with phonons) from the adiabatic minimum of the state has taken place. Perception of the observed photoluminescence spectra as being too narrow compared to expectations have been based on this assumption. By directly simulating spectra using a Jahn-Teller model, we investigate the dramatic effects possible, including spectral narrowing.

We also calculate emission lifetimes and competitive photochemical reaction rates. These include: rates for intersystem crossing reactions between the triplet and singlet manifolds, as well as rates for internal-conversion involving both avoided crossings (forming traditional transition states) and symmetry-forbidden processes at conical intersections. Only simple model calculations are reported, accurate to about an order of magnitude, but these are sufficient to capture processes occurring from sub-ps to ms or longer timescales that are relevant to the experimental observations and estimated photoemission lifetimes.

The results show how detailed knowledge of both the singlet and triplet manifolds, as well as the operation of the Jahn-Teller effect, is required to interpret the observed photoluminescence of $V_B^-$. In particular, three transitions, one in the singlet manifold and two in the triplet manifold, are pursued in detail concerning the likelihood that they could be responsible for the observed emission. We also measure improved defect spectra for comparison.

**Results**

**a. Measurement and interpretation of h-BN photoluminescence**

Figure 2 shows the original[5] photoemission spectrum observed from h-BN defects that display ODMR with a zero-field splitting parameter in the ground state $D_{gs}$ of 3.5 GHz, along with two newly recorded spectra. All new spectra are corrected for the response functions of the diffraction grating and detector, and were obtained at either 295 K or 77 K. As previously observed[5], such spectra show weak intensity, indicating low quantum yield, subsequent to photo-absorption. Also, the spectra arise from multiple emitters, distinct from the single-photon emitters commonly studied in h-BN[1-4]. The spectra are presented as the bandshape function $E(\nu)/\nu^3 \propto \lambda^5 E(\lambda)$ that display the intrinsic properties of the defect, allowing the shapes and bandwidths of photoemitters at different wavelengths to be directly compared. Maximum emission occurs at vertical transition energies $\Delta E_v^E$ of 1.5–1.6 eV (830–770 nm). The quantitative analysis of the effect of site inhomogeneity on the observed spectral bandshapes is not currently possible.

In SI, these and more spectra are presented and crudely analysed based on the (incorrect) assumption that each spectrum arises only from a single emitter. In this way, upper bounds can be determined for the reorganisation energy $\lambda^E$ associated with photoemission, as well as an estimate of the energies $E_{00}$ of the associated zero-phonon lines (ZPL). A traditional approach is taken in which each spectrum is represented using a set of Huang-Rhys factors depicting transitions between non-degenerate electronic states. This results in the spectral interpretation: $E_{00} = (1.61 \pm 0.03)$ eV and $\lambda^E < 0.05$ eV. Weaker emitters are apparent in the spectra, and in principle these could range in energy from 1.3 to 2.0 eV, based on the available information. The reorganisation energy could also be significantly smaller than the upper bound of 0.05 eV; indeed, the new spectra reported herein are significantly narrower than that originally reported (Figure 2), reflecting more the intrinsic properties of single defects. For comparison, traditional bright h-BN single-photon emitters have been categorised into "Group-1" emitters with reorganisation energies of 0.06–0.16 eV and "Group-2" emitters with 0.015–0.030 eV[11].

Two features of this analysis are important. First, the determined reorganisation energies do not include contributions from the acoustic phonons that determine the width of the ZPL, whereas the calculations presented later include all contributions. Second, the $V_B^-$ defect has inherent 3-fold symmetry and hence many of its electronic states will be doubly degenerate and therefore subject to the Jahn-Teller effect. In SI Section S2, we show that this effect can considerably reduce calculated spectral bandwidths, masking the effect of the reorganisation energy. Indeed, both effects can place the ZPL in the far blue-region of the observed spectral tail, as illustrated in Figure 2. As a result of both effects, the above traditional analysis could significantly underestimate the actual reorganisation energy.

**b. Overview of key orbital energy levels, electronic states, and the Jahn-Teller effect**

If a boron atom is removed from a pristine h-BN layer to make $V_B^-$, "dangling bonds" appear at each neighbouring nitrogen atom in both the σ and π electronic systems, making for six intrinsic orbital levels associated with the defect. We employ many different computational methods, all of which indicate that these six levels sit in the band gap of the h-BN, as sketched in Figure 3, that are occupied by 10 electrons, consistent with other calculations[12,17,18]. Varying occupancy, as depicted in Figure 4, results in many low-energy singlet and triplet defect states, most of which are open-shell in nature.

States always arise as mixtures of such configurations, but nevertheless we name them based on the configuration that is the most dominant; this labelling is therefore *diabatic* in nature and hence not subject to the discontinuities that occur at Jahn-Teller and other conical intersections, or the anharmonic effects that give rise to transition states.

We optimize the geometries of identified excited states, seeking potential-energy minima. Mostly, these optimizations are constrained to depict $C_{2v}$ symmetry; sometimes optimized geometries with $D_{3h}$ symmetry result, and sometimes vibrational analyses (or other means) indicate that the $C_{2v}$ structures are transition states rather than minima on the complete potential-energy surface. At $D_{3h}$ geometries, states are usually labelled using labels appropriate to $D_{3h}$, with structures at $C_{2v}$ geometries similarly named. State names based on these two labelling schemes are defined in Figure 4 for the configurations of greatest interest.

Four relevant, doubly degenerate states at $D_{3h}$ geometries are: $(1)^1E'$, $(1)^1E''$, $(1)^3E'$, and $(1)^3E''$. These must undergo Jahn-Teller distortions that reduce their symmetry to $C_{2v}$. Each component of a doubly degenerate state has the same label in $D_{3h}$ symmetry, but individual labels in $C_{2v}$, e.g., $(1)^3E''$ splits into $(1)^3A_2$ and $(1)^3B_1$, whilst $(1)^3E'$ splits into $(1)^3A_1$ and $(1)^3B_2$. Note that we use $C_{2v}$ standard axis conventions for planar molecules throughout[40,41] (otherwise "$B_1$" and "$B_2$" would be interchanged). Physically, this effect produces "tricorn Mexican hat" shaped potential-energy surfaces as a function of some $e'$ doubly degenerate vibrational coordinate, as sketched in Figure 5. A conical intersection appears at the central undistorted $D_{3h}$ geometry, with three equivalent local minima separated by three equivalent transition states, all of $C_{2v}$ symmetry, appearing on axes separated from each other by 120°. Away from these axes, the symmetry is reduced to $C_s$. The example shown in Figure 5 pertains to $(1)^1E'$; its components in $C_{2v}$ symmetry are $(1)^1A_1$ and $(1)^1B_2$, with calculations predicting that $(1)^1B_2$ forms the local minima (in this case, defining the singlet ground state), whilst $(1)^1A_1$ provides transition states that are unstable to distortion in a $b_2$ vibrational mode. As the figure shows, the three local minima support short N–N interaction distances within the defect, located along each of the three crystallographic axes, whilst the analogous distances are lengthened in the transition states. Note that the derivative discontinuity manifested at the conical intersection causes the symmetry to change abruptly as it is crossed along one of the three $C_{2v}$ orientations.

The basic tricorn Mexican hat depicted in Figure 5 may be distorted symmetrically, through avoided crossings or other interactions with nearby states, as well as distorted asymmetrically through strain effects. Although only briefly discussed herein, out-of-plane distortions may be introduced as well as the in-plane distortions manifested in the figure. The shortening of neighbouring N–N distances apparent in the figure highlights that defects rearrange their structure to maximise strong chemical bonding effects, and that this is state-dependent, for it relies on the electron occupancy and overall spin. Model compounds containing only a single ring (Figure 1) may over-accentuate this effect[24].

If the transition-state barrier displayed in Figure 5 is large compared to the associated vibrational energy spacings—or if distortions induced by interactions with other states or external forces are large—then just one single well of the tricorn structure needs to be considered in calculating spectroscopic and other properties, known as the *static* Jahn-Teller effect. Huang-Rhys-type models may provide realistic predictions of spectral bandshapes in this scenario. Alternatively, if the transition-state is of low energy, then quantum interference occurs between the vibronic energy levels in each of the three wells, possibly controlling spectral lineshapes; this is known as the *dynamic* Jahn-Teller effect. Note that all spectroscopic simulations reported herein pertaining to the Jahn-Teller effect are performed using a diabatic basis. They include both lower and upper adiabatic surfaces, with the Born-Oppenheimer approximation shown in Figure 5 never introduced.

**c. Electronic-structure calculations for the triplet manifold**

Key calculated properties of the triplet-state manifold for the $V_B^-$ defect of h-BN are presented in Table 3, evaluated for the model systems shown in Figure 1. Extended results, including more excited states up to 5.5 eV at vertical excitation, all ring sizes, various basis sets, and additional computational methods, are provided in SI Table S1–Table S4. Predictions made by the CAM-B3LYP, HSE06, MRCI, CCSD, CCSD(T), and EOMCCSD methods are mostly in good agreement, suggesting that the evaluated properties are reliable. Results from CASSCF are qualitatively similar but not of quantitative accuracy owing to its neglect of dynamics electron correlation, with CASPT2 results being unreliable owing to low-energy-denominator effects. The results presented for the 5-ring model obtained using both full CAM-B3LYP and that incorporated into a QM/MM scheme are very similar, with only the QM/MM scheme applied to 10- to 30-ring models. The presented results are: the vertical absorption energies $\Delta E_v^A$ at the ground-state optimised geometries, the adiabatic transition energies $\Delta E_0$ evaluated at individually optimised geometries, and the reorganisation energies associated with absorption ($\lambda^A$) and emission ($\lambda^E$). The adiabatic transition energies $\Delta E_0$ may be directly compared to observed ZPL energies $\Delta E_{00}$ by neglecting the (usually small, but at most 0.25 eV)[24] changes in zero-point energy $\Delta E_{zpt}$ that can occur. Also, the emission reorganisation energies $\lambda^E$ can be directly compared to the observed data in Figure 2 if non-degenerate states are assumed and allowed Franck-Condon intensity[42] rather than vibronically-borrowed Herzberg-Teller intensity[43] dominates.

That so many different methods predict similar results for the triplet manifold is unusual for defect spectroscopy[24]. The reason for this is that the ground state is predicted to be $(1)^3A_2'$, a state that contains seemingly full shells of both spin-up and spin-down electrons (Figure 4) and hence can be treated using conventional DFT and CC techniques; the important $(1)^3E''$ excited state has similar properties. As the ground state $(1)^3A_2'$ is well represented, TDDFT and EOMCCSD approaches are expected to provide good descriptions of all states that can be produced from it by single excitation. Prediction of this $(1)^3A_2'$ ground state agrees with previous calculations that are in accord with observed ODMR magnetic properties.[17,18]

Convergence of the CAM-B3LYP calculations with respect to expansion of the model compound from 1-ring to 30-ring compounds, extension to a 3-layer model, and further extension to include implicit treatment of the surrounding h-BN crystal, as well as basis-set expansion, is described in SI Section S3. In summary, the 1-ring model is only qualitatively indicative, the 2-ring model is adequate for most purposes, and the 3-ring model is quantitatively reliable. The correction needed to apply to 2-ring 6-31G* calculations to mimic 6-ring cc-pVTZ calculations in solid h-BN is (–0.01 ± 0.07) eV, and at most 0.15 eV in magnitude. Hence 2-ring 6-31G* calculations are identified as a computationally efficient approach of sufficient accuracy to support the comparison of calculated and observed data; later, mostly only calculations at this level are applied to the singlet manifold.

In $D_{3h}$ symmetry, the low-lying triplet excited states are predicted to be $(1)^3E''$ (forbidden Franck-Condon emission), $(1)^3A_1''$ (allowed Franck-Condon emission, lone lifetime, out-of-plane polarised), and $(1)^3E'$ (allowed Franck-Condon emission, short lifetime, in-plane polarised). Also, whilst $(1)^3E'$ is predicted to dominate absorption at the excitation wavelength used in the experiments (532 nm, 2.33 eV, see SI Table S16 and Figure S12), it appears to be too high in energy and its spectrum too broad to account for the emission process, so attention is focused onto the $(1)^3E''$ and $(1)^3A_1''$ states.

The $(1)^3E''$ first excited triplet state must undergo Jahn-Teller distortion, hence manifesting a tricorn Mexican hat potential-energy surface (see, e.g., Figure 5), leading to state components with $C_{2v}$ symmetry that are labelled $(1)^3A_2$ and $(1)^3B_1$. The relative ordering of these components is critical to understanding spectroscopic properties. CAM-B3LYP calculations on the 2-ring model compound predict that $(1)^3A_2$ is lower in energy by 0.18 eV, with correction to embed the defect in a 3D h-BN crystal changing this to 0.19 eV (Table 3). Similarly, EOMCCSD calculations predict a barrier of 0.15 eV. Also, CCSD/6-31G predicts a barrier of 0.43 eV, but this result is unreliable as this method incorrectly splits the energy of these state components by 0.23 eV at the $D_{3h}$ ground-state geometry. In contrast, MRCI predicts no barrier and CASSCF predicts that $(1)^3B_1$ is lower in energy by 0.49 eV. Other reported calculations for this splitting based on DFT and CASSCF approaches have also reported $(1)^3B_1$ as being of lower energy[17], but as these methods sometimes predict results similar to higher-level approaches and sometimes results that are very different, we are unable to estimate their reliability. The proper treatment of dynamical electron correlation is critical. We find that the MRCI Davidson correction is also significant. Hence the CAM-B3LYP and EOMCCSD results that embody more of the basic physics without inherent unreliability issues are taken as to be the most indicative. Note that the CAM-B3LYP and EOMCCSD calculations indicate that an avoided crossing between $(1)^3E''$ and $(1)A_1''$ upon $C_{2v}$ distortion is more significant than the Jahn-Teller effect and generates an extremely complex low-energy excited-state manifold, see e.g. SI Section S8, especially Figure S10 and Table S14.

Table 3 shows that most calculation methods predict that the adiabatic transition energy for $(1)^3A_2 \rightarrow (1)^3A_2'$ photoluminescence is in the range $\Delta E_0$ = 1.6–1.8 eV. This is in good agreement with the emission origin energies of $\Delta E_{00}$ ~ 1.6 eV (SI Figure S1) obtained using Huang-Rhys models to interpret the spectra as if they arise from single emission sources. Alternatively, the calculated emission reorganisation energies (Table 3) are $\lambda^E$ = 0.33 eV (CAM-B3LYP) and 0.27 eV (EOMCCSD) for the 2-ring model. The calculated correction to obtain CAM-B3LYP results for large rings with the cc-pVTZ basis embedded in 3D h-BN is –0.08 eV, reducing this to 0.25 eV. If the same correction is applied to the EOMCCSD results (Table 3), the value becomes 0.19 eV. As shown in Figure 2, the associated CAM-B3LYP spectrum is broader than that originally reported[5] and much broader than those reported herein. In SI Section S8, a Jahn-Teller spectral model is developed that simultaneously includes both the $(1)^3A_2$ and $(1)^3B_1$ components of the $(1)^3E''$ state. The best-estimate spectrum obtained from the CAM-B3LYP 2-ring model is shown in Figure 2. It displays apparent spectral narrowing and is much more reminiscent of the observed spectra. A crudely estimated EOMCCSD spectrum is also mooted in the figure, obtained by scaling the CAM-B3LYP spectrum by the estimated $\lambda^E$ difference, that is similar to our newly observed spectra. Note that, within the Jahn-Teller model, interchange of the ordering of $(1)^3A_2$ and $(1)^3B_1$ does not greatly perturb the spectral width as the critical lower-energy component always has the larger reorganisation energy (see SI Section S8).

It is possible that the CAM-B3LYP, EOMCCSD, and MRCI calculations misrepresents state ordering and that in fact $(1)^3A_1''$ lies lower than $(1)^3E'$. The calculations indicate that $(1)^3A_1''$ has a minimum of $D_{3h}$ symmetry plus also an associated set of $C_{2v}$ symmetry that we label $(2)^3A_2$ resulting from avoided crossings with higher-energy states, but the later appears to become unviable as ring-size increases. Considering the high-symmetry geometry, the calculated emission reorganisation energies (Table 3) for $(1)^3A_1'' \rightarrow (1)^3A_2'$ vary over the range 0.03–0.09 eV and are consistent with our new observed spectra, as demonstrated by the simulated spectrum shown in Figure 2.

**d. Electronic-structure calculations for the singlet manifold**

Based on the highly open-shell nature of the low-energy singlet states of $V_B^-$ as depicted in Figure 4, the methods properties listed in Table 2 indicate that standard particle-conserving single-reference methods, such as DFT, CCSD, or CCSD(T), should not provide qualitatively useful spectroscopic descriptions. The results presented in SI Section S6 indeed do indicate that this is the case. Alternatively, the particle-non-conserving DIP-EOMCC methodology is ideally-suited to this application, and mostly we focus on these results, seeking reliable error estimates for the calculated quantities. We also provide MRCI results, which, like those presented for the triplet manifold, are expected to be realistic, but nevertheless difficult to estimate error bounds for. All geometry optimisations are performed using CASSCF, as DFT methods are extremely unreliable for the singlet manifold. As for the triplet manifold, CASSCF typically predicts that the singlet states of interest are unstable to out of plane distortions, but likewise we also find that MRCI prefers high-symmetry structures instead. Even though there is no formal proof that MRCI (and more importantly DIP-EOMCC) predicts high symmetry structures, mostly, we confine discussion to their consideration only.

Details of the DIP-EOMCC calculations, emphasising convergence with respect to higher-order electron correlation effects and basis set size, are presented in SI Section S7. A full set of DIP-EOMCC results obtained using the 3-hole–1-particle $3h$-$1p$ approximation and a 6-31G basis set, abbreviated as DIP-EOMCC($3h$-$1p$)/6-31G or DIP($3h$-$1p$)/6-31G for short, is reported. Corrections to the raw DIP($3h$-$1p$)/6-31G data to include the leading high-order $4h$-$2p$ correlations outside the CCSD core, as well as the replacement of the 6-31G basis set by its larger 6-31G* counterpart, lead to extrapolated DIP($4h$-$2p$)/6-31G* values seen in Tables 4, S9, and S11. These corrections are mostly less than 0.1 eV in magnitude (see SI Section S7 for further details).

Calculated adiabatic transition energies and emission reorganisation energies within the singlet manifold are listed in Table 4, with vertical and adiabatic energy differences to $(1)^3A_2'$ listed in SI Table S11 and the vertical and adiabatic transition energies and reorganisation energies between all states considered listed in Table S10. The lowest-energy singlet state is predicted to be $(1)^1E'$, which undergoes a large Jahn-Teller distortion to form $(1)^1B_2$ minima and $(1)^1A_1$ transition states, as indicated in Figure 5. The next singlet state is predicted to be $(1)^1A_1''$, which distorts to $(1)^1A_2$. This is very close in energy to $(1)^1E''$, a state that undergoes a Jahn-Teller distortion to $(2)^1A_2$ (minima) and $(1)^1B_1$ (transition states), but the distortion is weak and the avoided crossing between $(1)^1A_2$ and $(2)^1A_2$ is important and taken to dominate the excited-state properties. A variety of states are apparent at energies ca. 0.5 eV higher, but we briefly consider only one of these, $(2)^1A_1$ owing to its close relationship to $(1)^1E'$ that is apparent from considering the diabatic state descriptions given in Figure 4. The properties of the singlet-state manifold are complex and best understood globally through the state-energy-minimum depiction provided in SI Figure S2.

Quantitatively, the lowest-energy singlet component $(1)^1B_2$ for the 2-ring model is predicted to lie adiabatically 0.56 eV above the triplet ground state $(1)^3A_2'$ (SI Table S11); applying the QM/MM procedure to expand the ring size to a 5-ring model (see SI Section S7) increases this to 0.84 eV. The lowest energy photoemission within the singlet manifold is similarly predicted to be $(1)^1A_2 \rightarrow (1)^1B_2$ at $\Delta E_0 = 1.25$ eV (Table 4). Next follows $(2)^1A_2 \rightarrow (1)^1B_2$ at $\Delta E_0 = 1.44$ eV, close to the observed emission energy. Of greatest note, the emission reorganisation energy for this is calculated to be $\lambda^E = 0.10$ eV, something possibly consistent with the very narrow observed photoluminescence spectra. That this transition could account for the observed photoluminescence peaks at 1.5–1.6 eV (Figure 2) therefore requires further consideration.

### e. State dipole moments: possible Stark shifts and long-range dielectric spectral shifts

The calculated dipole moment changes for the excited states of the $V_B^-$ of h-BN are described in SI Section S5. Even though these changes can be quite large and indicate substantial charge transfer within the inner ring of the defect, associated Stark effects are predicted to be small, with spectral shifts as large as 0.1 eV requiring nearby charges or ion pairs.

### f. Photoluminescence assignment possibilities.

The most likely origins of the observed photoluminescence analysed within the Huang-Rhys model to have $\Delta E_{00} \sim 1.6$ eV and $\lambda^E < 0.05$ eV is either the triplet transition $(1)^3E'' \rightarrow (1)^3A_2'$, with dominant $(1)^3A_2 \rightarrow (1)^3A_2'$ component for which the best calculations predict $\Delta E_0 = 1.78$–$1.83$ eV and $\lambda^E = 0.19$–$0.25$ eV, and/or the singlet transition $(1)^1E'' \rightarrow (1)^1E'$, with dominant $(2)^1A_2 \rightarrow (1)^1B_2$ component for which calculations predict $\Delta E_0 = 1.44$ eV and $\lambda^E = 0.10$ eV. Another possibility is also $(1)^3A_1'' \rightarrow (1)^3A_2'$ emission as this is predicted at $\Delta E_0 = 2.00$–$2.13$ eV and to be very narrow with $\lambda^E = 0.03$–$0.07$ eV.

If $(1)^3E'' \to (1)^3A_2'$ is responsible, then either the calculation methods all significantly misrepresent the reorganisation energy, or else the apparent spectral width must be narrower than what is trivially expected owing to the Jahn-Teller effect. If either $(1)^1E'' \to (1)^1E'$ or $(1)^3A_1'' \to (1)^3A_2'$ are involved, then the operative photochemical processes need to be of a type that would facilitate population buildup on the initial states for a sufficiently long period. This is in-principle possible as the observed quantum yield is very low.

In the following two subsections, to examine these possibilities, we consider sophisticated spectral simulation approaches followed by photochemical reaction-rate estimations.

### g. Spectral simulations

A variety of Huang-Rhys and Jahn-Teller spectral simulations are performed, as described in SI Section S8, with the principle results shown in Figure 2. Details including the form and displacement of the critical normal modes, their ring-size dependence, symmetry, contributions from Franck-Condon (allowed) and Herzberg-Teller (forbidden) intensity, and model dependences, are discussed therein. Full details including excited-state frequencies and the associated Duschinsky rotation matrices are also provided in SI data files. The absorption spectra predicted using individually determined vibrational modes for each excited state are also presented in SI Figure S12.

For the $(1)^3E'' \to (1)^3A_2'$ emission, a major result is that the Jahn-Teller effect is capable of narrowing apparent spectral widths, as shown in Figure 2, whilst maintaining the adiabatic transition energy $\Delta E_0$ and reorganisation energy $\lambda^E$. The Jahn-Teller effect withdraws intensity from the spectral wings to concentrate it around the vertical emission energy $\Delta E_v^E = \Delta E_0 - \lambda^E$. Note that, within the Jahn-Teller analysis, the ZPL is forbidden; all intensity is therefore associated with vibronic origins. As the Jahn-Teller distortion intrinsically permits allowed out-of-plane polarised emission, a vibronic origin results from the $e'$ distortion. In addition, the $(1)^3E''$ state may couple vibronically with the nearby $(1)^3E'$ state, facilitating additional vibronic origins with $e''$ symmetry. The intensity of such transitions has been calculated using Herzberg-Teller theory (see SI Table S15). These results indicate that the borrowed in-plane polarised intensity should be 7 times stronger than the intrinsic out-of-plane polarised contribution.

Concerning the possibility of $(1)^3A_1'' \to (1)^3A_2'$ emission, the predicted spectrum (Figure 2) is indeed very narrow and fully consistent with the observed narrow bandshape. Minimal Herzberg-Teller intensity is predicted for this transition, making it purely out-of-plane polarized. The spectral bandshape calculated for the $(2)^1A_2 \to (1)^1B_2$ (dominant) component of the $(1)^1E'' \to (1)^1E'$ emission is also in good agreement with experiments (Figure 2).

### h. Rates for photochemical reactions and photoemission

The low quantum yield of photoemission requires explanation. The rates of many photophysical and photochemical processes are determined from the calculated triplet and singlet potential energy surfaces in SI Section S9, with key results summarised in Figure 6 (77 K) and SI Figure S14 (295 K). Overviewing Figure 6, vertical absorption is predicted to be 2000 times stronger to $(1)^3E'$ than to $(1)^3A_1''$, with absorption to $(1)^3E'$ being Franck-Condon forbidden. After allowing for geometry relaxation, Jahn-Teller and avoided crossing interactions, and vibronic (Herzberg-Teller) borrowing, these ratios become 2000:1:8. Fast relaxation from $(1)^3E'$ quickly transfers the absorbed energy to either $(1)^3A_2$ or $(2)^3A_2$.

The $(1)^3E'' \to (1)^3A_2'$ photoemission is predicted to have a lifetime of 11 μs, much slower than the intersystem crossing to $(1)^1B_2$ which has a predicted lifetime of 3.8 ns at 77 K and 1.7 ns at 295 K. Calculated rates for this process are insensitive to details of the calculations such as the use of EOM-CCSD, or MRCI as the $(1)^1B_2$ surface crosses $(1)^3E''$ close to its $(1)^3A_2$ minimum. It is therefore a robust prediction of the calculations that intersystem crossing to the singlet manifold consumes most of the quantum yield, with the quantum yield for photoluminescence from within the triplet manifold being very low, ~0.03 %, and temperature insensitive, in qualitative agreement with the experimental data. The quantum yield is small but needs to be large enough to produce the observed ODMR contrast, which is suggestive of values of this order. Of the other possible scenarios considered for the photoemission, for the $(1)^3A_1'' \rightarrow (1)^3A_2'$ transition, the predicted emission lifetime, quantum yield, and temperature dependence are inconsistent with the experimental observations. For $(2)^1A_2 \rightarrow (1)^1B_2$, most predicted photochemical properties are highly inconsistent with those required. A feature of interest, however, is the long ground-rate recovery times of $10^{23}$ s at 77 K and 1.8 s at 295 K. These rates are sensitive to calculation details, with a 0.3 eV reduction in the calculated energy differences leading to times of 3 s and 2 μs, respectively. Hence the calculations cannot rule out the possibility that initial excitation converts most of the defects in the h-BN to their singlet state, with subsequent absorption and emission happening within the singlet manifold. Nevertheless, the photochemical data strongly suggests that the photoemission is $(1)^3E'' \rightarrow (1)^3A_2'$.

**Conclusions**

The reliable prediction of defect spectroscopic properties remains a severe challenge for both electronic-structure computation and spectral/photochemical simulation. Our conclusion for $V_B^-$ is that, to within likely errors in the calculations, only the $(1)^3E'' \rightarrow (1)^3A_2'$ emission is capable of explaining the observed emission associated with ODMR. This is despite all high-level computational methods used predicting the spectrum to be broader than those observed. Indeed, our new experimental measurements, presenting bands much narrower than previously observed, accentuate this effect. Issues such as the critical role of the Jahn-Teller effect in driving apparent spectral narrowing, and the role of acoustic phonons, demand further attention. Central to this is the loss of the generally accepted qualitative scenario that the ZPL is apparent in spectra, whereas our calculations perceive it as forbidden, but otherwise located in the far high-energy tail of the spectrum. One alternate assignment possibility cannot be eliminated, however, and that is that initial irradiation converts the defects into a long-lived singlet state, with subsequent absorption end emission pertaining to this manifold.

Concerning electronic-structure calculations, we see the need for reliable high-level methods with useful worst-case scenario error expectations. One of the key findings of this work is the demonstration of the ability of the DIP-EOMCC methodology to provide a reliable description of the complex singlet manifold of the $V_B^-$ defect in h-BN. Even basic DIP-EOMCC($3h$-$1p$)-level truncation opens up new possibilities for reliable modelling defects in h-BN and similar 2D materials. It could also be applied to less-difficult scenarios such as the triplet manifold of $V_B^-$. Associated with this is the coupling of such high-level methods with QM/MM schemes, allowing the QM part of this to accurately describe electronic effects and the MM part to simultaneously describe long-range nuclear structural effects. Such approaches should become the norm for defect spectroscopic modelling.

**Methods**

**a. Material fabrication**
The analysed samples were hBN flakes, neutron irradiated in the Triga Mark I IPR-R1 nuclear reactor (CDTN, Brazil), with a thermal flux of $4 \times 10^{12}$ n cm$^{-2}$ s$^{-1}$ for 16 h, with a resulting integrated dose of approximately $2.3 \times 10^{18}$ n cm$^{-2}$. All the samples were irradiated in cadmium capsules to block thermal neutrons and let the most energetic neutrons irradiate the sample[5].

**b. Spectroscopy measurements**
Spectroscopy measurements were carried out on a lab-built confocal microscope. A 532-nm, continuous-wave, solid-state laser (Gem 532TM; Laser Quantum Ltd.) was used as the excitation source. Light was focused onto the sample via a high numerical aperture (NA 0.9) air objective (Plan Fluor Epi P 100 ×; Nikon). Emission from the sample was collected in reflection, filtered through a long pass filter (transmission >560 nm) to suppress the excitation laser and sent into a multimode fibre. The collected signal could then be sent either to an avalanche photodiode (SPCM-AQRH-W4-FC; Excelitas Technologies) or a spectrometer with a 300 g/mm grating (SpectraPro Monochromator Acton SP2300), mounting a thermoelectric cooled (75 °C) CCD camera (Pixis Camera 256; Princeton Instruments). Spectra were acquired both at liquid nitrogen (77 K) and at room temperature (295 K).

**c. Electronic structure computations**

We utilise a wide range of computational methods, for which strengths and weaknesses are discussed in SI Section S1 and summarised in Table 2[25-30,34,35,44-73]. These methods include:

(1) CAM-B3LYP[60-62], as an example of an appropriate entry-level DFT methodology[24,73], as well as the commonly used HSE06 functional[58,59], both relying on the time-dependent formulation of DFT (TDDFT)[36] to determine excited electronic states. The D3(BJ) dispersion correction[74] is applied to all systems involving multiple h-BN layers.

(2) the standard single-reference coupled-cluster (CC) theory[64-66,68] with singly and doubly excited clusters (CCSD)[67] combined with a quasi-perturbative non-iterative correction due to connected triply excited clusters defining the widely used CCSD(T) approximation[69], along with the equation-of-motion (EOM) extension of CCSD to excited states abbreviated as EOMCCSD[35].

(3) The double-ionisation-potential (DIP) extension of the EOMCC formalism[25-30], abbreviated as DIP-EOMCC, using both the basic 3-hole–1-particle ($3h$-$1p$)[27-30] truncation and the highest currently implemented[29,30] $4h$-$2p$ level, which belong to a broader category of particle non-conserving EOMCC theories[29,30,68]. In the case of the $4h$-$2p$ truncation, we use active orbitals to select the dominant $4h$-$2p$ components to reduce computational costs[29,30]. These approaches allow one to determine singlet and triplet manifolds of open-shell systems that can formally be obtained by removing two electrons from the parent closed-shell cores (an operation generating the appropriate multi-configurational reference space within a single-reference framework), while relaxing the remaining electrons to capture dynamic electron correlations.

(4) The CASSCF approach[31-33], which is a conventional multi-reference technique for capturing static electron correlation effects, and

(5) Two different ways of correcting CASSCF calculations for dynamic correlations missing in CASSCF, including (CASPT2)[70,71], which uses the second-order multi-reference perturbation theory, and one of the variants of MRCI[34], which incorporates singly and doubly excited configuration state functions from a CASSCF reference, followed by the internal contraction and adding quasi-degenerate, relaxed-reference, Davidson corrections.

(6) Application of the ONIOM[75,76] approach to QM/MM to extend model sizes. This uses an AMBER[37] force field for the MM part, parameterised to mimic CAM-B3LYP/D3 results[38]. Two rings are retained in the QM part only, leading to very computationally efficient calculations.

All DFT and EOMCCSD calculations and CCSD geometry optimizations were performed using Gaussian-16[77]. All CASSCF, CASPT2, CCSD(T) and MRCI calculations were performed using MOLPRO[78]. All DIP-EOMCC calculations were carried out using standalone in-house codes[29,30] interfaced with the integral and SCF routines in GAMESS[79,80]. In the initial CC stages prior to DIP-EOMCC diagonalizations, these codes rely upon the spin-integrated CCSD routines available in GAMESS as well[81]. In all correlated calculations, the core orbitals correlating with the $1s$ shells of the B and N atoms were kept frozen and the spherical components of $d$ and $f$ basis functions, if present in the basis set, were employed throughout. The basis sets used were STO-3G[82], 6-31G[83], 6-31G*[83], and cc-pVTZ[84]. Self-consistent reaction field calculations, modelling a defect embedded deep within the h-BN bulk, are performed using the polarizable continuum model[85] using Gaussian-16, with the low-frequency bulk dielectric constant of h-BN is taken to be 5.87, whilst the high-frequency value is taken as 4.32.

### d. Observed spectral fitting

Observed spectra were fitted to a thermal Huang-Rhys model using the THRUP programme[86,87]. This allows spectra to be simulated for multiple electronic states interacting through multiple vibrational modes using either diabatic or adiabatic representations. It is often used to model the complex scenarios that arise during natural[88,89] and artificial[90] photosynthetic systems. In this application, it is used simply to model spectra within the Huang-Rhys model. Full details are given in SI Section S2.

### e. Spectral simulation

Absorption and emission spectra for assumed non-degenerate states are evaluated using Huang-Rhys-type schemes, perhaps extended to include curvilinear coordinates, Herzberg-Teller effects, and/or approximate inclusion of the Duschinsky matrix relating the initial-state and final-state normal modes using the DUSHIN software[91]. For systems of 10 rings or more, only the basic Huang-Rhys model is used, driven using analytical Hessian matrices written by Gaussian-16 into its formatted checkpoint files (use instead of the associated listed normal modes as their use was found to lead to significant errors). Methods beyond the above such Huang-Rhys-type approaches that include the Jahn-Teller effect are described in detail in SI. Sect S8.

### f. Photophysical and photochemical rate simulations

All methods used are traditional applications of either transition-state theory or adiabatic electron-transfer theory and are described in detail in SI Section S9[92-100].

**Data Availability**
All critical data concerning optimised structures and energies are provided in SI; anything else is available on request from the authors.

**Acknowledgments**
We thank the Australian Research Council (DP 160101301, DP190101058 and DE180100810), National Natural Science Foundation of China (NSFC; Grant 11674212), and the Chemical Sciences, Geosciences and Biosciences Division, Office of Basic Energy Sciences, Office of Science, U.S. Department of Energy (Grant No. DE-FG02-01ER15228) for funding this work and National Computational Infrastructure (grant no2) and Intersect (Grants d63, r88, sb4) as well as computer facilities at Shanghai and Michigan State Universities for computer time.

**Author contributions**
J.R.R. and M.J.F. conceived the project, whilst P.P. devised the computational strategy involving the DIP-EOMCC methodology and I.A. the new experimental strategies. J.S. performed the required DIP-EOMCC calculations, C.B. and M.K the measurements, and J.R.R. carried out the DFT, MRCI, and CCSD computations, spectral simulations and rate modelling. All authors contributed to manuscript production, led by J.R.R.

**Table 1.** Key aspects of h-BN defects and their consequences for high-level spectroscopic modelling[24].

| Aspects of h-BN defects | Consequences |
| --- | --- |
| Mostly open-shell in character | Static electron correlation often critical, posing problems for DFT and CC calculations that focus on single-reference configurations, demanding MRCI or, when appropriate, TDDFT, EOMCCSD, EOM-CC etc. methods. |
| Dynamically correlated | Methods such as Hartree-Fock theory, CASSCF (and DMRG) give poor results, establishing approaches such as DFT, CC, QMC, and MRCI as entry-level methods. |
| Strong electron-hole interactions | The ordering of electronic states can be very different to that suggested by considering only one-particle orbital energy levels, demanding extensive state searching. |
| Electronic arrangements modulate chemical bonding | Very large structural rearrangements can accompany electronic transitions, often leading to very large reorganisation energies, e.g., 2.5 eV, so that state energy ordering at adiabatically relaxed geometries can be very different to that perceived vertically at the ground-state geometry. |
| Charge transfer can occur | DFT methods such as PBE and HSE06 can, without warning, deliver very poor results, identifying range-corrected functionals such as CAM-B3LYP[63] as the entry level for DFT calculations. |
| Are embedded in 3D materials | Dielectric effects can be critical, but as h-BN is essentially a 2D material, such effects are minimised[101]. |

**Table 2.** The electronic-structure computational methods used, their key properties and applicability.[a]

| Method | Properties | Applicability |
| --- | --- | --- |
| DFT | includes dynamic electron correlation but fails for open-shell systems | the critical triplet ground state $(1)^3A_2'$ |
| TDDFT | excited-state open-shell systems are well described if the ground-state is closed shell | triplet excited states e.g., $(1)^3E''$, $(1)^3A_1''$, $(1)^3E'$, including depiction of the normal vibrational modes of phonons |
| CCSD | very good for closed-shell systems, treats static electron correlation asymmetrically | the critical triplet ground state $(1)^3A_2'$, perhaps other triplet states |
| CCSD(T) | typically improves on CCSD using perturbative corrections for triple excitations, but often degrades performance for open-shell systems | the critical triplet ground state $(1)^3A_2'$ |
| DIP-EOMCC | works well for open-shell systems for which a suitable closed-shell reference is available containing two additional electrons | all states of interest, but no analytical gradients for geometry optimisation |
| CASSCF[b] | good description of static electron correlation, poor description of dynamic electron correlation[c] | all states qualitatively described correctly, poor quantitative accuracy and therefore below entry level |
| CASPT2 | CASSCF plus perturbative treatment of dynamic electron correlation[c] | all states, may give poor results if low-energy coupled states are nearby |
| MRCI | CASSCF plus treatment of dynamic electron correlation up to double excitations, but size inconsistent[c] | all lowest-energy states of each spin and spatial symmetry, no analytical gradients for geometry optimisation |

a: see SI Section S1 for discussion and explanations.
b: Enhanced treatment of static electron correlation using large active spaces is warranted and can be achieved using density-matrix renormalisation group (DMRG) approaches, and can also be empirically parameterised using DFT orbitals; these features have been applied[17] to $V_B^-$ but are not incorporated herein.
c: All approaches based on CASSCF suffer from the limitations of the need to choose an active space and the possible use of state averaging. The active space used herein is shown in Figure 2; with state averaging used where possible to establish overall symmetry and energy relativity. Even though calculation errors may be relatively low, these issues make error bounds difficult to estimate as it is impractical to demonstrate convergence.

**Table 3.** Calculated spectroscopic properties for absorption and emission transitions within the triplet manifold of the $V_B^-$ defect in h-BN, in eV, involving the $(1)^3A_2'$ triplet ground state.

| METHOD | CAM-B3LYP | QM/MM | CAM-B3LYP | CASSCF[f] | CASPT2[b] | MRCI[b] | CCSD | CCSD(T)[d] | EOM-CCSD[c] | EOM-CCSD | Other[k] |
|---|---|---|---|---|---|---|---|---|---|---|---|
| BASIS | cc-pVTZ[a] | 6-31G* | 6-31G* | 6-31G* | 6-31G* | 6-31G* | 6-31G | 6-31G | 6-31G* | cc-pVTZ[a] | |
| RINGS | 6[a] | 30 | 2 | 2 | 2 | 2 | 2 | 2 | 2 | 6[a] | 2D |
| LAYERS | crystal[a] | 1 | 1 | 1 | 1 | 1 | 1 | 1 | 1 | crystal[a] | 1 |
| | | | | Vertical absorption energies $\Delta E_v^A$ | | | | | | | |
| $(1)^3E'' - (1)^3A_2$ | 2.04 | | 1.99 | 1.45[ehi] | 1.84[o] | 1.98[eh] | 2.18[m] | 2.12 | 2.16 | 2.21 | 1.92 |
| $(1)^3E'' - (1)^3B_1$ | 2.04 | | 1.99 | 1.37[ehi] | | 1.93[ehi] | 2.41 | 2.25 | 2.16 | 2.21 | 1.92 |
| $(1)^3A_1''$ | 2.03 | | 2.08 | 2.25[io] | 2.17[io] | 2.13[o] | | | 2.27 | 2.22 | |
| $(1)^3E' - (1)^3A_1$ | 2.65 | | 2.76 | 3.01[h] | 2.94[h] | 2.98[h] | 2.82 | 2.70 | 2.76 | 2.66 | 2.29 |
| $(1)^3E' - (1)^3B_2$ | 2.65 | | 2.76 | | | | | | 2.76 | 2.66 | 2.29 |
| $(1)^3A_2''$ | | | 3.70 | | | | | | 3.93 | | 1.8 |
| | | | | Adiabatic transition energies $\Delta E_0$ | | | | | | | |
| $(1)^3A_2$ | 1.78[g] | 1.74 | 1.67 | 1.34[h] | 1.83[h] | 1.51[h] | 1.60[c] | | 1.72 | 1.83 | > 1.72 |
| $(1)^3B_1$[j] | 1.97 | | 1.85 | 0.85[hi] | | 1.51[h] | 2.03[c] | | 1.87 | 1.99 | 1.72 |
| $(1)^3A_1''$ | 2.00 | | 2.04 | 2.18[n] | 1.54[n] | 2.12[n] | | | 2.17 | 2.13 | |
| $(2)^3A_2$ | 1.99 | | 1.92 | 2.02[io] | 2.08[io] | 2.18[o] | | | 1.94 | 2.01 | |
| $(1)^3A_1$[l] | 2.22 | | 2.21 | 2.32 | 2.33 | 2.37 | 2.06[c] | | 2.19 | 2.20 | |
| $(2)^3B_2$[l] | 2.25 | | 2.21 | | | | | | 2.22 | 2.26 | |
| | | | | Absorption reorganisation energies $\lambda^A$ | | | | | | | |
| $(1)^3A_2$ | 0.26 | | 0.32 | 0.11 | | 0.47 | | | 0.44 | 0.37 | |
| $(1)^3B_1$ | 0.08 | | 0.14 | 0.52 | | 0.42 | | | 0.29 | 0.22 | 0.20 |
| $(1)^3A_1''$ | 0.03 | | 0.04 | | | | | | 0.09 | 0.07 | |
| $(2)^3A_2$ | 0.04 | | 0.16 | 0.24 | 0.09 | | | | 0.33 | 0.20 | |
| $(1)^3A_1$ | 0.43 | | 0.55 | 0.69 | 0.61 | 0.61 | | | 0.57 | 0.45 | |
| $(2)^3B_2$ | 0.40 | | 0.55 | | | | | | 0.54 | 0.39 | |
| | | | | Emission reorganisation energies $\lambda^E$ | | | | | | | |
| $(1)^3A_2$ | 0.25 | 0.26 | 0.33 | 0.20[h] | 0.49[h] | 0.31[h] | | | 0.27 | 0.19 | |
| $(1)^3B_1$ | 0.12 | | 0.13 | -0.42[h] | | 0.20[h] | | | 0.11 | 0.10 | |
| $(1)^3A_1''$ | 0.06 | | 0.05 | | | | | | 0.02 | 0.03 | |
| $(2)^3A_2$ | 0.15 | | 0.12 | | | | | | 0.09 | 0.12 | |
| $(1)^3A_1$ | 0.48 | | 0.42 | 0.41[h] | 0.52[h] | 0.51[h] | | | 0.44 | 0.50 | |
| $(2)^3B_2$ | 0.46 | | 0.41 | | | | | | 0.47 | 0.52 | |

a: After addition of CAM-B3LYP corrections to mimic calculations on the h-BN crystal for the cc-pVTZ basis set, see SI Table S3.
b: At CASSCF optimised geometries.
c: CCSD/6-31G at CAM-B3LYP/6-31G* optimised geometries.
d: At CCSD optimised geometries.
e: CASSCF calculations can break the degeneracy of degeneracy of $E'$ and $E''$ states owing to asymmetric representation of the active space and orbital optimisation. This effect can be minimised using state-averaged approaches, and is usually reduced by MRCI.
f: CASSCF predicts symmetry lowering of $(1)^3A_2'$ to (at least) $C_{2v}$, influencing $\lambda^E$, but this symmetry lowering is not supported by single-point energy MRCI calculations; CCSD can also predict (much smaller) distortions not supported at the CCSD(T) level.
g: Leads to a ZPL energy of $\Delta E_0 = 1.76$ eV after addition of the calculated zero-point energy correction of $\Delta E_{zpl} = -0.02$ eV obtained for the 2-ring compound using CAM-B3LYP/6-31G*.
h: single-state calculation.
i: two-state calculation using 50:50 weighting; both *h* and *i* flags indicate that each approach gives this result.
j: transition state on the tricorn Mexican hat, see e.g. Figure 5.
k: By Ivády et al.[17]
l: CAM-B3LYP 2-ring calculations indicate $(1)^3E'$ undergoes barrierless out-of-plane relaxation from $(1)^3A_1$ to $(1)^3A_2$ and from $(2)^3B_2$ to $(2)^3A_2$.
m: 2.02 eV by CCSD/6-31G*.
n: at CAM-B3LYP/6-31G* geometry.

o: three-state calculations weighted 40:30:30 predict (CASSCF, CASPT2, MRCI) energies $\Delta E_v^A$= (2.17, 1.63, 2.13) eV and $\Delta E_0$ = (2.22, 1.67, 2.18) eV; the CASPT2 results, which predict that $(1)^3A_1''$ is the lowest-energy triplet excited state, are believed unreliable.

**Table 4.** Calculated spectroscopic properties for emission transitions within the singlet manifold of the $V_B^-$ defect in h-BN, in eV, involving the $(1)^1E'$ lowest-energy singlet state[a] of the 2-ring model compound and that as embedded in a 5-ring model.

| initial state | to $(1)^1B_2$ | | | | | to $(1)^1A_1$ | | | |
|---|---|---|---|---|---|---|---|---|---|
| | MRCI | DIP(3h-1p)[b] | DIP(4h-2p)[b] | QM/MM[c] DIP(4h-2p)[b] | Other[d] | MRCI | DIP(3h-1p)[b] | DIP(4h-2p)[b] | QM/MM[c] DIP(4h-2p)[b] |
| | | | | Adiabatic transition energies $\Delta E_0$ | | | | | |
| $(1)^1A_2$ | 1.15 | 1.01 | 0.98 | 1.25 | | 0.98 | 0.94 | 0.86 | 1.09 |
| $(2)^1A_2$ | 1.18 | 1.31 | 1.35 | 1.44 | 1.33 | 1.01 | 1.24 | 1.23 | 1.28 |
| $(1)^1B_1$ | 1.23 | 1.35 | 1.41 | 1.50 | | 1.06 | 1.28 | 1.29 | 1.34 |
| | | | | Emission reorganisation energies $\lambda^E$ | | | | | |
| $(1)^1A_2$ | 0.29 | 0.24 | | | | | 0.19[e] | | |
| $(2)^1A_2$ | 0.26 | 0.10[e] | | | | 0.15 | 0.03 | | |
| $(1)^1B_1$ | 0.26 | 0.10[e] | | | | 0.15 | 0.03 | | |

a: the minimum of the lowest-energy singlet state $(1)^1E'$ is predicted to be its $(1)^1B_2$ component, 0.4 - 0.7 eV higher in energy than the triplet ground state $(1)^3A_2'$ (see SI Table S9).
b: $\Delta E_0$ values calculated using the raw DIP(3h-1p)/6-31G and extrapolated DIP(4h-2p)/6-31G* data (see SI Table S9).
c: 5-ring total, 2-ring in QM part, frozen 2-ring structure with the outer 3 rings optimised.
d: by Ivády et al.[17]
e: to lower Born-Oppenheimer surface, as is relevant to high-energy emission; for data pertinent to the native upper Born-Oppenheimer surface, see SI Table S11 and Figure S2 as they provide broader perspectives.

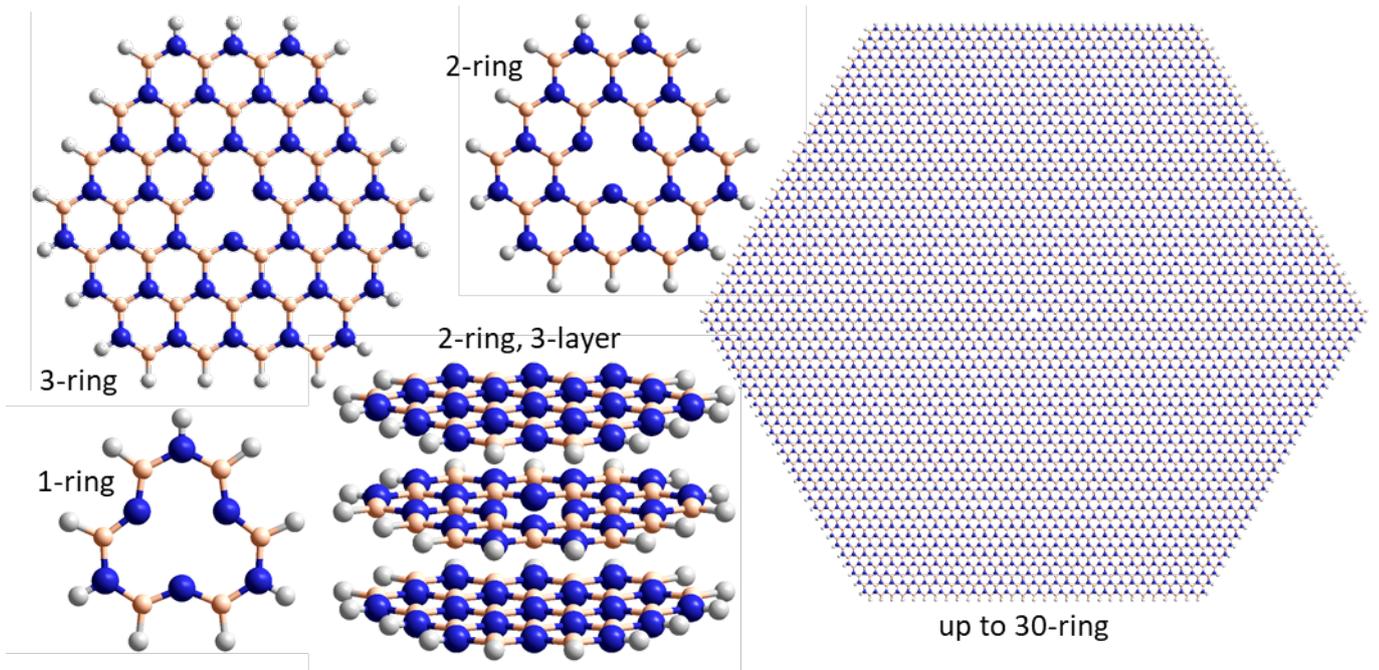

**Figure 1.** Model compounds used to study the electronic states of the $V_B^-$ defect: one layer- 1-, 2-, 3- and 30-ring shown (also 4-, 5-, 6-, 10-, 15-, 20- and 25-ring models not shown), as well as a 2-ring 3-layer model and that plus a dielectric continuum to model bulk h-BN. Blue, peach, and grey spheres represent nitrogen, boron, and hydrogen atoms, respectively.

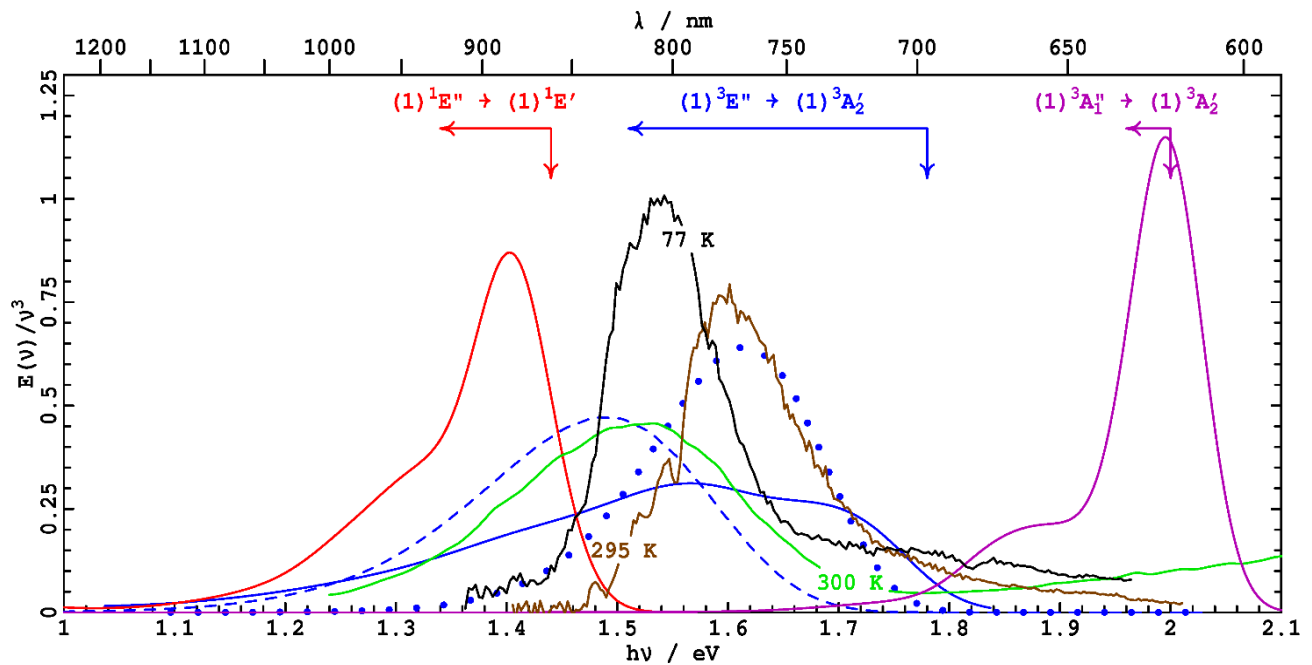

**Figure 2.** Observed ensemble photoemission bandshapes of h-BN defects reported[5] as displaying ODMR with ground state zero field splitting $D_{gs} = 3.5$ GHz are compared to calculated bandshapes for three possible emissions. Observed spectra: at 295 K (brown) and 77 K (black) from current measurements, and at 300 K from previous ones[5] (green). Calculated spectra are for the transitions $(1)^1E'' \rightarrow (1)^1E'$ (red), $(1)^3E'' \rightarrow (1)^3A_2'$ (blue), and $(1)^3A_1'' \rightarrow (1)^3A_2'$ (purple), obtained using DIP-EOMCC Huang-Rhys (red), CAM-B3LYP Huang-Rhys (purple and blue solid), CAM-B3LYP Jahn-Teller (blue dashed), and CAM-B3LYP Jahn-Teller crudely adjusted to match the EOMCCSD $\Delta E_0$ and $\lambda^E$ (blue dots). Arrows indicate CAM-B3LYP or DIP-EOMCC ZPE locations $\Delta E_0$ and spectral widths $\lambda^E$.

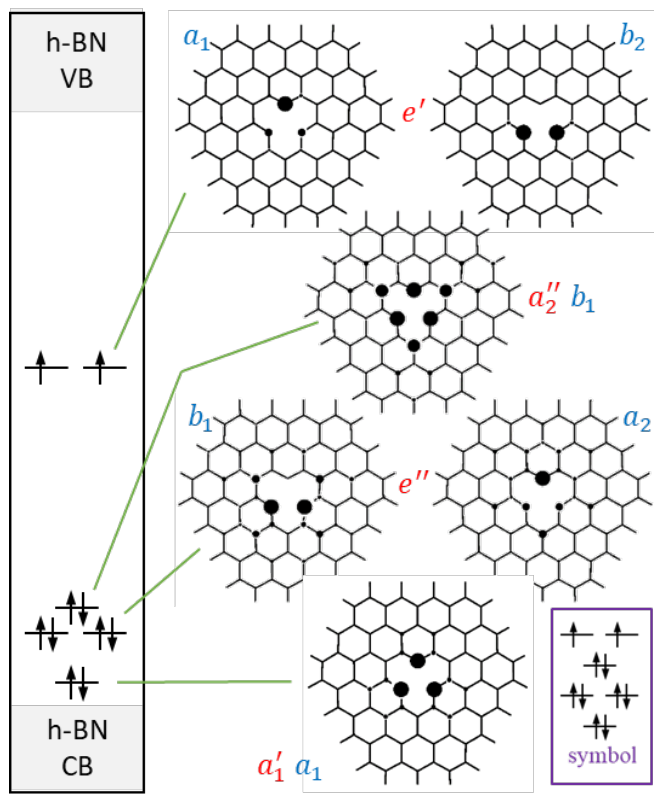

**Figure 3.** The six 3-ring CAM-B3LYP/6-31G* mid-gap defect orbital energy levels (spin restricted) lying between the h-BN valence band (VB) and conduction band (CB), represented as circles depicting the atomic electron density, for the $V_B^-$ defect in h-BN. Symmetries are indicated for both the $D_{3h}$ (red) and $C_{2v}$ (blue) point groups. The electronic configuration of the $(1)^3 A_2'$ ground state is shown, with excited states depicted in Figure 4 using variants of the inserted symbol.

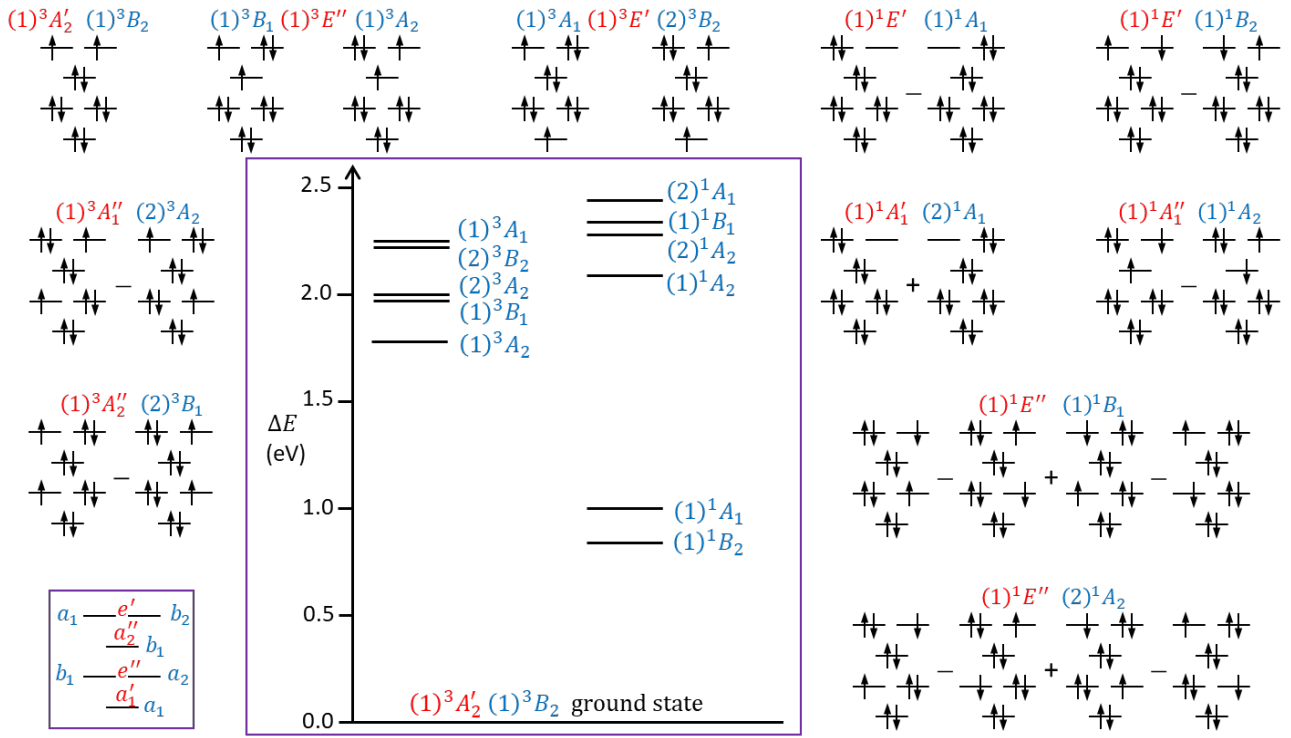

**Figure 4.** Key diabatic configurations of orbital energy levels contributing to the low-energy states of the $V_B^-$ defect in h-BN, showing symmetry labels depicting both $D_{3h}$ (red) and $C_{2v}$ (blue) local point-group symmetry. Adiabatic wavefunctions obtained from the electronic structure calculations are depicted throughout the text in terms of their dominant diabatic configurations, with often considerable mixing apparent that is method dependent. The inset shows the best-estimate adiabatic energy minima: triplets EOMCCSD, see Table 3, singlets: DIP-EOMCC except $(2)^1A_1$ from MRCI, see Table S11; (*nb.*, singlet-triplet splittings appear to be overestimated).

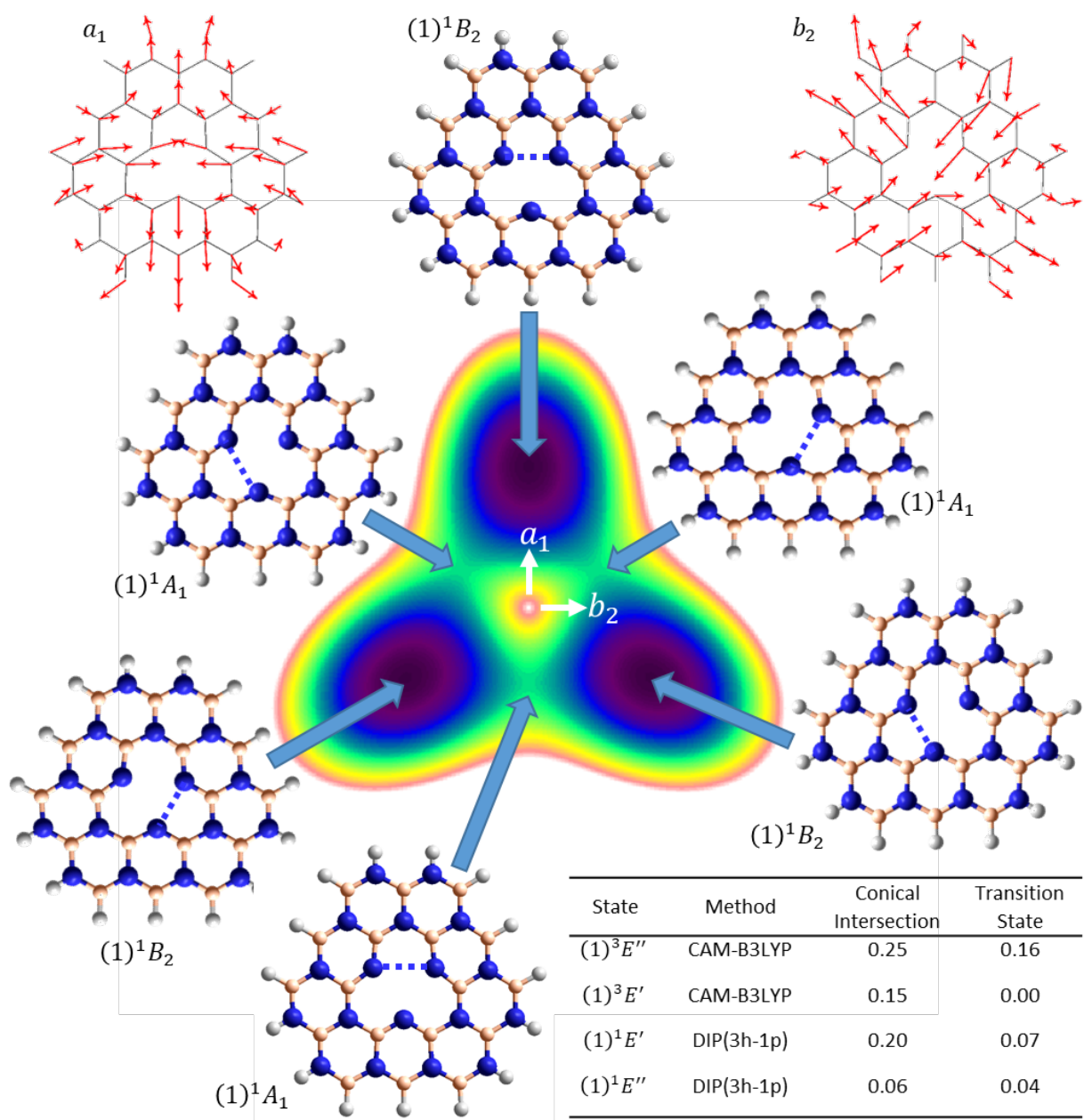

**Figure 5.** Contour plot (black- low energy minima, white- energy of the conical intersection and above) depicting generic tricorn Mexican-hat Born-Oppenheimer potential-energy surface associated with Jahn-Teller conical intersections. The energy is shown as a function of displacements in the $a_1$ and $b_2$ components of some generalised $e'$ vibrational mode, with indicated the ground-state normal-mode displacement vectors for the dominant mode involved in most transitions, $34e'$ at 183 cm$^{-1}$ (0.023 eV) varying to 163–192 cm$^{-1}$ in the triplet excited states. The conical intersection in the centre has $D_{3h}$ local-point group symmetry, whereas the three lines passing through the stationary points have $C_{2v}$ symmetry and all other points have $C_s$ symmetry. The six indicated chemical structures were optimised using CASSCF(10,6)/6-31G* for the $(1)^1E'$ surface and depict $(1)^1B_2$ local-minima (black) and $(1)^1A_1$ transition-state (cyan) structures for the $(1)^1E'$ surface. The dashed lines indicate B-B separations that differ from those involving the analogous inner-ring B atom. The key shows the calculated energies of the conical intersections and transition states with respect to the minima for different $V_B^-$ states, in eV. For $(1)^3E'$ and $(1)^1E'$, avoided crossings at $C_{2v}$ geometries with $(1)^3A_1''$ and $(1)^1A_1''$, respectively, significantly distort the Jahn-Teller surfaces. For more details, see SI Figure S3 - Figure S7 and Figure S11, as well as Table S12 - Table S14, and Table S17.

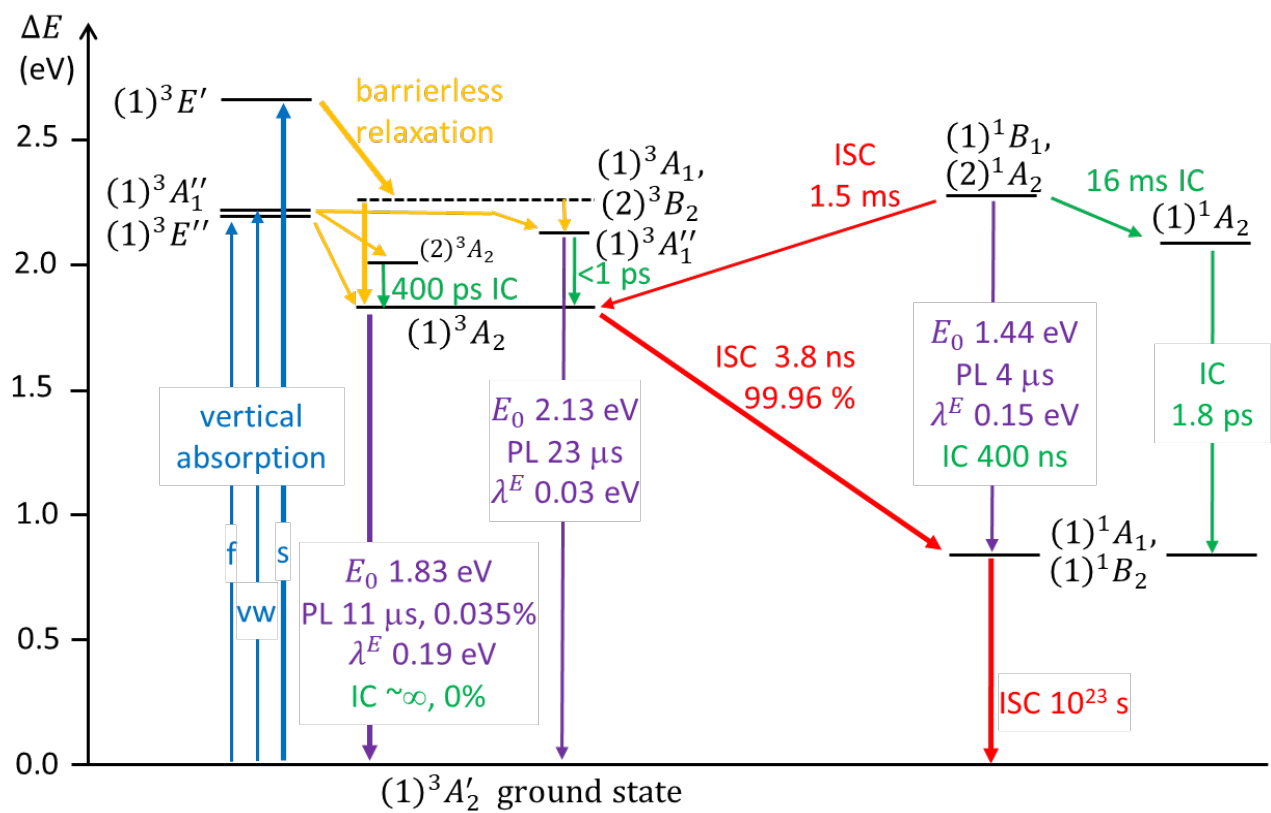

**Figure 6.** Calculated photochemical and photoemission processes for the $V_B^-$ defect of h-BN at 77 K, obtained using EOMCCSD calculations on the triplet manifold, DIP-EOMCC calculations on the singlet manifold, and MRCI-calculated transition dipole moments and spin-orbit couplings. Thicker arrows indicate the critically perceived processes during the photocycle. The indicated processes are: blue- vertical absorption (related absorption at lower energies down to the shown ZPLs and also at higher energies will also occur, see SI Fig. S12), with Franck-Condon (Herzberg-Teller) allowed oscillator strengths in the ratio f:vw:s of 0(8):1:2000; orange- barrierless ultrafast relaxation to ZPLs; green- internal conversion (IC); red- intersystem crossing (ISC), purple- photoluminescence (PL). Marked percentages indicate quantum yields. The energy levels and internal rate processes the doubly degenerate states that form transition states in $C_{2v}$ symmetry are not shown, for clarity (see instead Figure 4). States denoted with a solid line indicate that a local minima is established (triplet manifold) or believed (singlet manifold) to be involved, dashed lines indicate saddle structures that are unstable to out-of-plane distortion leading directly to $(1)^3A_2$. The use of different computational methods for the singlet and triplet manifolds results in overestimation of the singlet-triplet splittings. Some rates like the primary 3.8 ns ISC are insensitive to calculation details and temperature, whereas others re extremely sensitive to both, *e.g.,* the shown $10^{23}$ s singlet recovery time can be reduced to the seconds timescale within possible computational uncertainties. Analogous results for 295 K are shown in SI Figure S14.